\documentclass[11pt]{article}
\usepackage{graphicx}
\usepackage{cite}

\begin{document}

%
% Title, Authors, Affiliations
% ===================
%
\title{\Large\bf
Distribution of linearly polarized gluons inside a large nucleus}

\author{Andreas Metz, Jian Zhou
 \\[0.3cm]
{\normalsize\it Department of Physics, Barton Hall,
  Temple University, Philadelphia, PA 19122, USA}}

\maketitle

% headline
%\thispagestyle{fancy}
%\fancyhead[R]{\tt \jobname}

%
% Abstract
% ======
%
\begin{abstract}
\noindent
The distribution of linearly polarized gluons inside a large nucleus is studied
in the framework of the color glass condensate.
We find that the Weizs\"acker-Williams distribution saturates the positivity bound
at large transverse momenta and is suppressed at small transverse momenta,
whereas the dipole distribution saturates the bound for any value of the transverse
momentum.
We also discuss processes in which both distributions of linearly polarized gluons
can be probed.
\end{abstract}

%
% 1. Section: Introduction
% ==================
%
\section{Introduction}
\noindent
Recently, transverse momentum dependent parton distributions (TMDs)~\cite{Collins:1981uw}
inside a nucleon have attracted a lot of interest.
Since TMDs not only depend on the longitudinal momentum fraction $x$ of the parton but
also on its transverse momentum $k_{\perp}$, they contain more detailed information on
the internal structure of the nucleon as compared to the conventional collinear
parton distributions.
So far, the main focus of the field has been on quark TMDs.
In particular, the (na\"ive) time-reversal odd quark Sivers function~\cite{Sivers:1989cc}
and Boer-Mulders function\cite{Boer:1997nt}, which are intimately linked to
initial/final state interactions of the active quark~\cite{Brodsky:2002cx,Collins:2002kn},
have been under intensive investigation.
In comparison, the available studies of (polarized) gluon
TMDs~\cite{Mulders:2000sh,Ji:2005nu,Anselmino:2005sh,Meissner:2007rx,Boer:2009nc,Boer:2010zf,
Qiu:2011ai} are still rather sparse.
Among them the distribution of linearly polarized gluons inside an unpolarized nucleon
($h_{1}^{\perp g}$ in the notation of Ref.~\cite{Meissner:2007rx}) is of particular
interest.
It is the only polarization dependent gluon TMD for an unpolarized nucleon, and therefore
may be considered as the counterpart of the quark Boer-Mulders function.
However, in contrast to the latter, $h_1^{\perp g}$ is time-reversal even implying that
initial/final state interactions are not needed for its existence.
It has been shown that this distribution, in principle, can be accessed through measuring,
e.g., azimuthal $\cos 2\phi$ asymmetries in processes such as jet or heavy quark pair
production in electron-nucleon scattering as well as nucleon-nucleon scattering, and photon
pair production in hadronic collisions~\cite{Boer:2009nc,Boer:2010zf,Qiu:2011ai}.
Such measurements should be feasible at RHIC, the LHC, and a potential future Electron
Ion Collider (EIC)~\cite{Anselmino:2011ay,Boer:2011fh}.

It has long been recognized that the $k_\perp$ dependent unpolarized gluon distribution
$f_1^g$ (also frequently referred to as the unintegrated gluon distribution) plays a central
role in small $x$ saturation phenomena.
Due to the presence of a semi-hard scale (the so-called saturation scale), generated
dynamically in high energy scattering, $f_1^g(x,k_{\perp})$ at small $x$ can be computed
using an effective theory which is also known as the color glass condensate (CGC)
framework (see~\cite{McLerran:1993ni,Mueller:2001fv,Iancu:2002xk,JalilianMarian:2005jf,Gelis:2010nm}
and references therein).
There are two widely used $k_{\perp}$ dependent unpolarized gluon distributions with
different gauge link structures: (1) the Weizs\"acker-Williams (WW)
distribution~\cite{McLerran:1993ni,Kovchegov:1996ty,JalilianMarian:1996xn}, and (2)
the so-called dipole distribution which appears, for instance, in the description of
inclusive particle production in $pA$ collisions~\cite{Dumitru:2001jn,Gelis:2002ki}
(see, e.g., Refs.~\cite{JalilianMarian:2005jf,Gelis:2010nm} for an overview).
The WW distribution describes the gluon number density and as such has a probability
interpretation, whereas the dipole distribution is defined as the Fourier transform of
the color dipole cross section.
Very recent work has demonstrated that both types of $k_\perp$ dependent gluon
distributions can be directly probed through two-particle correlations in various high energy
scattering reactions~\cite{Dominguez:2010xd,Dominguez:2011wm}.
These studies make use of an effective TMD factorization at small $x$ in the correlation
limit, where the transverse momentum imbalance of, e.g., two outgoing jets is much smaller
than the individual transverse jet momenta.
Such a factorization is suggested by the CGC approach.

In this Letter, we extend the calculation of $f_1^g(x,k_{\perp})$ to the case of $h_1^{\perp g}(x,k_{\perp})$.
To be more specific, we compute both the WW distribution and the dipole
distribution of linearly polarized gluons in the CGC framework. It
is shown that the WW distribution saturates the positivity
bound~\cite{Mulders:2000sh} at high transverse momenta, and is
suppressed at low transverse momenta. The dipole distribution
saturates the bound for any value of $k_\perp$. Following the
procedure outlined in~\cite{Dominguez:2010xd,Dominguez:2011wm} we
further argue that the WW distribution and the dipole distribution
can be accessed by measuring a $\cos 2\phi$ asymmetry for dijet
production in lepton nucleus scattering and for production of a
virtual photon plus a jet in nucleon nucleus scattering,
respectively. In some sense this also extends related
studies~\cite{Boer:2009nc,Boer:2010zf,Qiu:2011ai} to the small $x$
region.

%
% 2. Section: Weizs\"acker-Williams distribution
% ===========================
%
\section{Weizs\"acker-Williams distribution}
\noindent
We start the derivation by introducing the operator definition of the WW gluon distribution
inside a large nucleus~\cite{Mulders:2000sh,Meissner:2007rx},
\begin{eqnarray} \label{e:ww}
M_{WW}^{ij} & = & \int \frac{d \xi^- d^2 \xi_{\perp}}{(2\pi)^3 P^+} \,
e^{ixP^+\xi^- - i\vec{k}_{\perp} \cdot \vec{\xi}_\perp}
\langle A | F^{+i}(\xi^- + y^-, \xi_{\perp} + y_{\perp}) \, L_{\xi+y}^\dag  \, L_{y} \,
F^{+j}(y^-,y_\perp) |A \rangle
\nonumber\\ & = &
\frac{\delta_{\perp}^{ij}}{2} \, x f_{1,WW}^{g}(x, k_\perp) +
\bigg(\frac{1}{2}\hat k_\perp^i \hat k_\perp^j - \frac{1}{4} \delta_{\perp}^{ij} \bigg)
x h^{\perp g}_{1,WW}(x, k_\perp) \,,
\end{eqnarray}
where $\hat{k}_\perp^i = k_\perp^i/k_{\perp}$ ($k_{\perp} \equiv |\vec{k}_{\perp}|$).
Color gauge invariance is ensured by two (future-pointing) gauge links in the adjoint
representation.
We use
\begin{equation}
L_{\xi} = \mathcal{P} \, e^{-ig \int_{\xi^-}^{\infty^-} d \zeta^- A^+(\zeta^-, \xi_{\perp})} \,
\mathcal{P} \, e^{-ig \int^\infty_{\xi_{\perp}}
d\vec{\zeta}_{\perp} \cdot \vec{A}_\perp(\zeta_{\perp},\xi^-=\infty^-)} \,,
\end{equation}
where $A^\mu=A_a^\mu t_a$ with $(t_a)_{bc} = -i f_{abc}$, and $f_{abc}$ denoting the structure
constants of the $SU(3)$ group.
When the gauge links become unity by choosing the light-cone gauge with advanced boundary
condition, the above two distributions have a number density interpretation.
Note that our convention for $h_1^{\perp g}$ differs from the previous
literature~\cite{Mulders:2000sh,Meissner:2007rx} by a factor $k_\perp^2/M^2$, where $M$ is the
target mass.
We also would like to mention that the violation of translational invariance inside a large
nucleus prevents us from shifting the coordinate $y$ to zero, which would lead to the
conventional form of $k_\perp$ dependent gluon distributions.
However, as shown in Ref.~\cite{Dominguez:2011wm}, the gluon distributions defined in
Eq.~(\ref{e:ww}) are the ones which can be directly related to physical observables.

We perform the calculation of the WW gluon distributions in the CGC framework in the
light-cone gauge by following the standard procedure (see, e.g., Ref.\cite{Iancu:2002xk}
for an overview).
By solving the classical Yang-Mills equation of motion with a random color source, which
has only a plus component, the gluon field strength tensor reads
\begin{eqnarray}
F^{+i}(y_\perp) = \partial^+ A^i(y_\perp) =
- U(y_\perp) \partial_{\perp}^i \alpha U^\dag(y_\perp) \,.
\end{eqnarray}
All nonlinear effects are encoded in the Wilson line
$U^\dag(y_\perp)= \mathcal{P} \exp\{ig \int_{y^-}^\infty d \zeta^- \alpha (\zeta^-, y_\perp)\}$.
The quantity $\alpha$ satisfies the equation
$-\nabla_\perp^2 \alpha_a(y_\perp)=\rho_a(y_\perp)$ with $\rho_a$ being the color source
in covariant gauge.
We proceed by inserting this expression into the matrix element in~(\ref{e:ww}) and
by contracting the field operators in all possible ways.
Due to rotational symmetry and the ordering of the Wilson lines in the minus direction, the
only allowed contraction is
\begin{eqnarray}
\Big\langle F^{+i}(\xi+y) F^{+j}(y) \Big\rangle_A & = &
\Big\langle [U^\dag_{ab} \partial_{\perp}^i \alpha_b](\xi+y)
              [U^\dag_{ac} \partial_{\perp}^j \alpha_c](y) \Big\rangle_A
\nonumber \\
& = & \Big\langle \partial_{\perp}^i \alpha_b(\xi+y) \partial_{\perp}^j \alpha_c(y) \Big\rangle_A
\Big\langle U^\dag_{ab}(\xi+y) U_{ca}(y) \Big\rangle_A
\nonumber \\
& = &\delta(\xi^-) \Big\langle \mathrm{Tr} \, U^\dag (\xi+y) U(y) \Big\rangle_A
\Big[ -\partial_\perp^i \partial_\perp^j \Gamma_A( \xi_\perp) \Big] \lambda_A(y^-) \,,
\end{eqnarray}
where $U^\dag_{ac} = U_{ca}$ in the adjoint representation.
In the last step, we use the propagator
$\langle \alpha_a(x) \alpha_b(y) \rangle_A =
\delta_{ab} \, \delta(x^- - y^-) \, \Gamma_A( x_\perp-y_\perp) \, \lambda_A(x^-)$
with $\Gamma_A( k_\perp) = 1/k_{\perp}^4$, where $\lambda_A$ comes from the correlation of
color sources generated by a Gaussian weight function
$W_A[\rho]= \exp \Big\{-\frac{1}{2} \int d^3 x
\frac{ \rho_a(x) \rho_a(x)}{\lambda_A(x^-)} \Big\}$\cite{McLerran:1993ni},
\begin{equation}
\Big\langle \rho_a(x) \rho_b(y) \Big\rangle_A =
\delta_{ab} \, \delta^2(x_\perp-y_\perp) \, \delta(x^- - y^-) \, \lambda_A(x^-) \,.
\end{equation}
One can further evaluate the contraction of two Wilson lines by expanding them in
powers of $\alpha$, which leads to
\begin{equation}
\Big \langle \mathrm{Tr} \, U^\dag(\xi+y) U(y) \Big \rangle_A =
(N_c^2-1) \, \exp \bigg\{ -g^2 N_c [\Gamma_A(0_\perp) -\Gamma_A(\xi_\perp)]
\int_{y^-}^{\infty^-} d\zeta^- \, \lambda_A(\zeta^-) \bigg\} \,.
\end{equation}
Collecting all the pieces one obtains
\begin{eqnarray}
M_{WW}^{ij} & = & \frac{N_c^2-1}{4 \pi^3} \int dy^- d^2 y_\perp d^2 \xi_\perp \,
e^{-i\vec{k}_\perp \cdot \vec{\xi}_\perp}
\Big[ -\partial_\perp^i \partial_\perp^j \Gamma_A(\xi_\perp) \Big] \lambda_A(y^-)
\nonumber \\
&& \hspace{1.5cm} \mbox{} \times
\exp \bigg\{ -g^2 N_c [\Gamma_A(0_\perp) -\Gamma_A(\xi_\perp)]
\int_{y^-}^{\infty^-} d\zeta^- \, \lambda_A(\zeta^-) \bigg\} \,.
\end{eqnarray}
Integrating out $y_\perp$ and $y^-$ we end up with
\begin{equation}
M_{WW}^{ij} = \frac{N_c^2-1 }{4 \pi^3} \, S_\perp \int d^2 \xi_\perp \,
e^{-i\vec{k}_\perp \cdot \vec \xi_\perp}
\frac{ -\partial_\perp^i \partial_\perp^j \Gamma_A(\xi_\perp)}
{\frac{1}{4 \mu_A} \xi_\perp^2 Q_s^2}
\bigg( 1 - e^{ - \frac{\xi_\perp^2 Q_s^2}{4}} \bigg) \,,
\end{equation}
where $S_\perp= \pi R_A^2$ is the transverse area of the target nucleus,
$\mu_A = \int_{-\infty^-}^{\infty^-} dy^- \lambda_A(y^-)$,
and $Q_s^2 = \alpha_s N_c \mu_A{\rm ln} \frac{1}{\xi_\perp^2 \Lambda_{QCD}^2}$ is the saturation
scale.
By appropriate projections one can now obtain both TMD gluon distributions in a large nucleus.
For the unpolarized distribution one has
\begin{eqnarray}
x f_{1,WW}^g(x,k_\perp) & = & \delta_{\perp}^{ij} M_{WW}^{ij}
\nonumber\\
& = & \frac{N_c^2-1 }{N_c} \frac{S_\perp }{4 \pi^4 \alpha_s} \int d^2 \xi_\perp \,
e^{-i \vec{k}_\perp \cdot \vec{\xi}_\perp} \,
\frac{1}{\xi_\perp^2} \bigg( 1 - e^{ - \frac{\xi_\perp^2 Q_s^2}{4}} \bigg) \,,
\end{eqnarray}
in full agreement with already existing
calculations~\cite{Kovchegov:1996ty,JalilianMarian:1996xn}.
For the distribution of linearly polarized gluons we find
\begin{eqnarray} \label{e:res_h1perp_ww}
x h^{\perp g}_{1,WW}(x,k_\perp) & = &
\Big( 4\hat k_\perp^i \hat k_\perp^j - 2 \delta_{\perp}^{ij} \Big) M_{WW}^{ij}
\nonumber\\&=&
\frac{N_c^2-1 }{4 \pi^3} \, S_\perp \int d \xi_\perp \,
\frac{K_2 (k_\perp \xi_\perp)}{\frac{1}{4 \mu_A} \xi_{\perp} Q_s^2}
\bigg( 1 - e^{ - \frac{\xi_\perp^2 Q_s^2}{4}} \bigg) \,.
\end{eqnarray}
To arrive at the result in~(\ref{e:res_h1perp_ww}) we made use of the Bessel functions
$K_\nu(x)=\frac{i^\nu}{2 \pi} \int_{-\pi}^{\pi} d\theta \, \exp\{ix \cos \theta+i\nu \theta\}$,
and the recursion relation
$\frac{d}{dx} \Big[ \frac{K_\nu(x)}{x^\nu} \Big] = - \frac{ K_{\nu+1}(x)}{x^\nu}$.
Note that both $f_{1,WW}^g$ and $h^{\perp g}_{1,WW}$ depend, in particular, also on the
CGC parameter $Q_s$.

Let us now discuss the expression in Eq.~(\ref{e:res_h1perp_ww}) in the limit of high and low
transverse momenta.
For $k_{\perp} \gg Q_s$, the integral is dominated by small distances
$\xi_\perp \ll 1/Q_s$ and can be evaluated by expanding the exponential
$\exp \Big\{ - \frac{\xi_\perp^2 Q_s^2}{4}\Big\}$, leading to
\begin{equation}
x h^{\perp g}_{1,WW}(x,k_{\perp}) \; \simeq \;
2S_\perp \frac{N_c^2-1}{4 \pi^3}\frac{\mu_A}{k_\perp^2}
\qquad (k_{\perp} \gg Q_s) \,.
\end{equation}
For $\Lambda_{QCD} \ll k_{\perp} \ll Q_s$ the dominant contribution comes from large
distances $\xi_{\perp} \gg 1/Q_s$, where one can neglect the exponential.
We further neglect the logarithmic $\xi_{\perp}$-dependence in the saturation scale $Q_s$
and arrive at
\begin{equation}
x h^{\perp g}_{1,WW}(x,k_{\perp}) \; \simeq \; 2S_\perp \frac{N_c^2-1}{4 \pi^3}\frac{\mu_A}{Q_s^2}
\qquad (\Lambda_{QCD} \ll k_{\perp} \ll Q_s) \,.
\end{equation}
On the other hand, in these limits the unpolarized gluon distribution takes the form
\begin{eqnarray}
x f_{1,WW}^g(x,k_{\perp}) & \simeq &
S_\perp \frac{N_c^2-1}{4 \pi^3}\frac{\mu_A}{k_\perp^2}
\qquad (k_{\perp} \gg Q_s) \,,
\\
x f_{1,WW}^g(x,k_{\perp}) & \simeq &
S_\perp \frac{N_c^2-1}{4 \pi^3}\frac{1}{\alpha_s N_c} \,
\mathrm{ln} \frac{Q_s^2 }{k_\perp^2}
\qquad (\Lambda_{QCD} \ll k_{\perp} \ll Q_s) \,.
\end{eqnarray}
From those results one immediately finds that for large $k_{\perp}$
the distribution of linearly polarized gluons saturates the
positivity limit, which in our notation reads $h_1^{\perp g} \le 2
f_1^g$~\cite{Mulders:2000sh}.
This is actually not a very surprising result because, like for the unpolarized gluon
distribution, the correct perturbative tail~\cite{Nadolsky:2007ba,Catani:2010pd,Metz_Zhou:prep}
can be recovered for $h_1^{\perp g}$ at large $k_\perp$, for which one also finds complete
linear polarization.
In contrast, the ratio $h_{1,WW}^{\perp g}/f_{1,WW}^g$ is suppressed in the region of
small $k_\perp$, where gluon re-scattering effects play a more important role.

%
% 3. Section: Dipole distribution
% ===========================
%
\section{Dipole distribution}
\noindent
We now proceed to the calculation of the dipole distribution.
In that case the operator definition
reads~\cite{Bomhof:2006dp,Dominguez:2010xd,Dominguez:2011wm}
\begin{eqnarray}
M_{DP}^{ij} & = & 2 \int \frac{d \xi^- d^2 \xi_\perp}{(2\pi)^3 P^+} \,
e^{ixP^+\xi^- - i\vec{k}_{\perp} \cdot \vec{\xi}_{\perp}}
\langle A | \mathrm{Tr} \, F^{+i}(\xi^- + y^-, \xi_{\perp} + y_{\perp})  U_{\xi+y}^{[-]\dag }
F^{+j}(y^-,y_\perp) U_{\xi+y}^{[+]} |A \rangle
\nonumber\\ & = &
\frac{\delta_{\perp}^{ij}}{2} \, x f_{1,DP}^{g}(x, k_\perp) +
\bigg(\frac{1}{2}\hat k_\perp^i \hat k_\perp^j - \frac{1}{4} \delta_{\perp}^{ij} \bigg)
x h^{\perp g}_{1,DP}(x, k_\perp) \,,
\end{eqnarray}
where $U^{[-]}_\xi=U^n(0,-\infty; 0)U^n(-\infty,\xi^-; \xi_\perp)$
and $U^{[+]}_\xi=U^n(0,+\infty; 0)U^n(+\infty,\xi^-; \xi_\perp)$
are gauge links in the fundamental representation.
In covariant gauge, the only nontrivial component of the field strength tensor is
$F^{+i}(y_\perp) = -\partial_{\perp}^i \alpha(y_{\perp})$, which can be viewed as
the realization of the eikonal approximation in the McLerran-Venugopalan model.
By noticing this fact, one may easily see that
$U^n[+\infty, \xi^-; \xi_\perp] F^{+i}(\xi) U^{n \dag}[-\infty, \xi^-; \xi_\perp] \propto
\partial_\perp^i U^{n \dag}[-\infty, +\infty; \xi_\perp]$~\cite{Dominguez:2011wm}.
This relation finally leads to
\begin{equation}
M_{DP}^{ij}=\frac{k_\perp^i k_\perp^j N_c}{2 \pi^2 \alpha_s} S_\perp
\int \frac{d^2 \xi_{\perp}}{(2\pi)^2} \, e^{-i\vec k_\perp \cdot \vec \xi_\perp} \,
e^{-\frac{Q_{sq}^2 \xi_\perp^2}{4}} \,,
\end{equation}
where $Q_{sq}^2 = \alpha_s C_F \mu_A{\rm ln} \frac{1}{\xi_\perp^2 \Lambda_{QCD}^2}$ is the
quark saturation momentum.
Contracting $M_{DP}^{ij}$ with the different tensors one readily finds
\begin{equation}
x h^{\perp g}_{1,DP}(x, k_\perp) = 2 x f_{1,DP}^g(x, k_\perp)=
\frac{k_\perp^2N_c}{ \pi^2 \alpha_s} S_\perp
\int \frac{d^2 \xi_{\perp}}{(2\pi)^2} \, e^{-i\vec{k}_\perp \cdot \vec{\xi}_\perp} \,
e^{-\frac{Q_{sq}^2 \xi_\perp^2}{4}} \,,
\label{e:DP_bound}
\end{equation}
which means that the positivity bound is saturated for any value of $k_{\perp}$.
Note that both $f_{1,DP}^g$ and $h^{\perp g}_{1,DP}$ depend, in particular, also on the
CGC parameter $Q_{sq}$.
At large $k_\perp$, the correct perturbative tails are recovered for both the WW and the
DP distributions in the unpolarized and the polarized case, while the DP-type distributions
are more suppressed than the WW-type distributions at small $k_\perp$.
For more discussion and additional physical insights about the difference between
the two type distributions, see~\cite{Gelis:2010nm,Dominguez:2011wm} and references
therein.

%
% 4. Section: Observables
% =======================================
%
\section{Observables}
\noindent
The extraction of gluon TMDs through two-particle correlations in various high energy scattering
processes at small $x$ relies on an effective TMD factorization valid in the correlation
limit~\cite{Dominguez:2010xd,Dominguez:2011wm}, where the transverse momentum imbalance between
two final state particles (or jets) is much smaller than the individual transverse momenta.
Normally, higher twist contributions at small $x$ are equally important as the leading twist
contribution because of the high gluon density.
Therefore, in order to arrive at the mentioned effective TMD factorization, an analysis including
all higher twist contributions would be crucial.
For the unpolarized case it has been shown that the results from the effective TMD factorization
are in agreement with the results obtained by extrapolating the CGC calculation to the correlation
limit~\cite{Dominguez:2010xd,Dominguez:2011wm}.
By applying a corresponding power counting in the correlation limit, we find a complete matching
between the effective TMD factorization and the CGC calculation in the polarized case as well.
For simplicity, we will express our results in terms of $k_\perp$ dependent gluon distributions
rather than multi-point correlation functions.

First, we discuss dijet production in lepton nucleus scattering.
In fact, we consider the process $ \gamma^{\ast} + A \to q(p_1)+ \bar q(p_2) + X$ for
both transversely and longitudinally polarized photons.
(We also keep the quark mass $m_q$ in the calculation.)
Because there are only final state interactions between the $q \bar{q}$ pair and the target
nucleus, the correct gluon TMD entering the factorization formula is the WW distribution, in
which the final state interactions, to all orders, are resummed in future-pointing Wilson lines.
The calculation provides
\begin{eqnarray}
\frac{d \sigma^{\gamma^*_T A \to q \bar{q} + X}}{d P.S.} & = &
\delta(x_{\gamma^*}-1) H_{\gamma^*_T g \to q \bar{q}}
\bigg\{x f_{1,WW}^g(x, k_\perp)
\nonumber \\
&& - \frac{[z_q^2+(1-z_q)^2]\epsilon_f^2 P_\perp^2-m_q^2 P_\perp^2}
 {[z_q^2+(1-z_q)^2](\epsilon_f^4+ P_\perp^4)+2m_q^2 P_\perp^2}
 \cos (2 \phi) x h^{\perp g}_{1,WW}(x, k_\perp) \bigg\} \,,
\\
\frac{d \sigma^{\gamma^*_L A \to q \bar{q} + X}}{d P.S.} & = &
\delta(x_{\gamma^*}-1) H_{\gamma^*_L g \to q \bar{q}}
\bigg\{x f_{1,WW}^g(x, k_\perp)+
\frac{1}{2} \cos (2 \phi)  x h^{\perp g}_{1,WW}(x, k_\perp) \bigg\} \,,
\end{eqnarray}
where $x_{\gamma^*} = z_q+z_{\bar q}$, with $z_q$, $z_{\bar q}$ being the momentum fractions
of the virtual photon carried by the quark and antiquark, respectively.
The phase space factor is defined as $d P.S.=dy_1 dy_2 d^2P_\perp d^2 k_\perp$, where $y_1$,
$y_2$ are rapidities of the two outgoing quarks in the lab frame.
Moreover, $\vec P_\perp=(\vec p_{1\perp}-\vec p_{2\perp})/2$, and
$\epsilon_f^2=z_q(1-z_q)Q^2+m_q^2$.
The transverse momenta are defined in the $\gamma^{\ast} A$ {\it cm} frame.
In the correlation limit, one has
$|P_\perp| \simeq |p_{1\perp}| \simeq |p_{2\perp}| \gg |k_\perp|=|p_{1\perp}+p_{2\perp}|$.
The (azimuthal) angle between $\vec{k}_{\perp}$ and $\vec{P}_{\perp}$ is denoted by $\phi$.
The hard partonic cross sections $H_{\gamma^*_{T,L} g \to q \bar{q}}$ can be found in
Ref.~\cite{Dominguez:2011wm}.
Our calculation for these coefficients agrees with the results of~\cite{Dominguez:2011wm}.
The $\cos(2\phi)$-modulation of the cross section allows one to address the distribution
of linearly polarized gluons.
For intermediate values of $x$ this was already pointed out in Ref.~\cite{Boer:2010zf}.
Our calculation indicates that the largest azimuthal asymmetry can be expected for
longitudinally polarized photons.

Let us now turn to the dipole distribution at small $x$.
From a theoretical point of view, the simplest process to address $h_{1,DP}^{\perp g}$
seems to be back-to-back virtual photon plus jet production in $pA$ collisions, i.e.,
$p + A \to \gamma^{\ast}(p_1) + q(p_2)+X$.
For this reaction, the multiple gluon attachments to the initial and final state quark
line can be resummed which leads to the dipole type gluon distribution.
Moreover, in the forward rapidity region of the proton, one may simplify the calculation
by adopting an hybrid strategy~\cite{Gelis:2002ki} in which the dense target nucleus is
treated as color glass condensate, while on the side of the dilute projectile proton one
uses ordinary integrated parton distributions.
Even though, to the best of our knowledge, a general proof of this method is still missing
we use it here for the process under discussion.
The differential cross section, obtained in the effective TMD factorization, reads
\begin{eqnarray}
\frac{d \sigma^{p A \to \gamma^{\ast} q + X}}{d P.S.}
& = & \sum_q x_p f_1^q(x_p) \bigg\{ H_{qg \to \gamma^{\ast}q} \, x f_{1,DP}^g(x,k_{\perp})
+ \cos (2 \phi)H_{qg \to \gamma^{\ast}q}^{\cos (2 \phi)} \, x h_{1,DP}^{\perp g}(x,k_{\perp})
\bigg\}
\nonumber \\
& = & \sum_q x_p f_1^q(x_p) \, x f_{1,DP}^g(x,k_{\perp}) H_{qg \to \gamma^{\ast}q}
\bigg\{ 1 + \cos (2 \phi) \frac{2 Q^2 \hat{t}}{\hat{s}^2 + \hat{u}^2 + 2 Q^2 \hat{t}}
\bigg\} \,,
\label{e:cs_gamma_jet}
\end{eqnarray}
where the partonic cross sections are given by
\begin{equation}
H_{qg \to \gamma^{\ast}q} = \frac{\alpha_s \alpha_{em} e_q^2}{N_c \hat s^2}
\bigg( -\frac{\hat{s}}{\hat{u}} - \frac{\hat{u}}{\hat{s}}
- \frac{2Q^2 \hat{t}}{\hat{s} \hat{u}} \bigg) \,, \qquad
H_{qg \to \gamma^{\ast}q}^{\cos (2 \phi)} = \frac{\alpha_s \alpha_{em} e_q^2}{N_c \hat s^2}
\bigg( \frac{-Q^2 \hat{t}}{\hat{s} \hat {u}} \bigg) \,.
\end{equation}
Here we used the partonic Mandelstam variables $\hat{s} = (p_1 + p_2)^2$, $\hat{u} = (p_1 - p)^2$
and $\hat{t} = (p_2 - p)^2$, with $p$ denoting the momentum carried by the incoming quark from
the proton.
Note that the $\cos 2\phi$ modulation drops out for prompt (real) photon production.
In the second line in Eq.~(\ref{e:cs_gamma_jet}) we made use of the
relation~(\ref{e:DP_bound}), implying that the azimuthal dependence of the cross section is
completely determined by kinematical factors.

%
% 5. Section: Summary
% ================
%
\section{Summary}
\noindent
We derived both the WW distribution and the dipole distribution of linearly polarized gluons
in a large nucleus by using the CGC formalism.
The WW distribution saturates the positivity bound at large values of $k_{\perp}$, while
it is power-suppressed (compared to the unpolarized distribution) in the region of low
$k_{\perp}$.
The dipole distribution saturates the bound for any value of $k_{\perp}$.
It is worthwhile to point out that in the quark target model, treated to lowest nontrivial
order in perturbation theory, $h_1^{\perp g}$ also saturates the bound for small $x$ and
large $k_{\perp}$~\cite{Meissner:2007rx}.
We also computed $h_{1}^{\perp g}$ at large $k_\perp$ in the intermediate $x$ region
within standard collinear twist-2 factorization~\cite{Metz_Zhou:prep}.
Taking the dominant contribution at small $x$ again leads to saturation of the positivity
limit.
It would be interesting to explore how the distribution of linearly polarized gluons
behaves under QCD evolution effects.
We aim at addressing this important issue in future work.

We further argued that the WW and the dipole gluon distribution can be probed by measuring
a $\cos 2 \phi$ asymmetry for dijet production in DIS, and for virtual photon-jet production
in $pA$ collisions, respectively.
Such observables can, in principle, be measured at a future Electron Ion Collider, at RHIC
and the LHC.
Studying such effects could open a new path in spin physics.
Moreover, the results for the asymmetries constitute parameter-free predictions of the CGC
framework.
Therefore, exploring these observables may open complementary ways to test this effective
theory of small $x$ physics.
\\

%
% Acknowledgments
% ===============
%
\noindent
{\bf Acknowledgments:}
This work has been supported by the NSF under Grant No.~PHY-0855501.

%
% References
% =========
%

\end{document}